\documentclass[prb,showpacs,print,twocolumn]{revtex4}
\usepackage{amssymb}
\usepackage{mathrsfs}
\usepackage{graphicx}
\usepackage{dcolumn}
\usepackage{bm}
\usepackage{txfonts}

\begin{document}

\title{Spin-dependent transport for armchair-edge graphene
nanoribbons between ferromagnetic leads}

\author{Benhu Zhou,$^1$ Xiongwen Chen,$^1$ Benliang Zhou,$^1$ Kai-He
Ding,$^2$}
\author{Guanghui Zhou$^{1,3}$}
\email{ghzhou@hunnu.edu.cn}

\affiliation{$^1$Department of Physics and Key Laboratory for
Low-Dimensional Structures and Quantum Manipulation (Ministry of
Education), Hunan Normal University, Changsha 410081, China}

\affiliation{$^2$Department of Physics and Electronic Science,
Changsha University of Science and Technology, Changsha 410076,
China}

\affiliation{$^3$International Center for Materials Physics, Chinese
Academy of Sciences, Shenyang 110015, China}

\begin{abstract}
We theoretically investigate the spin-dependent transport for the
system of an armchair-edge graphene nanoribbon (AGNR) between two
ferromagnetic (FM) leads with arbitrary polarization directions at
low temperatures, where a magnetic insulator is deposited on the
AGNR to induce an exchange splitting between spin-up and -down
carriers. By using the standard nonequilibrium Green's function
(NGF) technique, it is demonstrated that, the spin-resolved
transport property for the system depends sensitively on both the
width of AGNR and the polarization strength of FM leads. The
tunneling magnetoresistance (TMR) around zero bias voltage possesses
a pronounced plateau structure for system with semiconducting 7-AGNR
or metallic 8-AGNR in the absence of exchange splitting, but this
plateau structure for 8-AGNR system is remarkably broader than that
for 7-AGNR one. Interestingly, the increase of exchange splitting
$\Delta$ suppresses the amplitude of the structure for 7-AGNR
system. However, the TMR is enhanced much for 8-AGNR system under
the bias amplitude comparable to splitting strength. Further, the
current-induced spin transfer torque (STT) for 7-AGNR system is
systematically larger than that for 8-AGNR one. The findings here
suggest the design of GNR-based spintronic devices by using a
metallic AGNR, but it is more favorable to fabricate a
current-controlled magnetic memory element by using a semiconducting
AGNR.

\end{abstract}

 \vspace{0.2cm} \maketitle

\section{Introduction}
Spintronics utilizes the electron spin degree of freedom to carry
information in electronic devices and functionalities. Its central
task is to control electron spin for information storage and
precessing.$^{1}$ Graphene, a single layer of carbon, has low
intrinsic spin-orbit and hyperfine couplings,$^{2}$ long spin
diffusion lengths ($\sim$2$\mu$ m).$^{3}$ These features suggest
that graphene is a promising candidate for the potential
applications in spintronics. Triggered by these findings, a large
number of works have been devoted to the spin-dependent transport in
graphene and graphene-FM heterostructures, as prototypical
spintronic devices. For example, a graphene spin-valve device has
been successfully fabricated showing that spin-polarized currents
can be injected by means of magnetic (cobalt, permalloy, etc.)
electrodes,$^{4-9}$ which shows high spin-polarized injection
efficiency. Meanwhile, a 10\% magnetoresistance (MR) is observed
$^{10}$ in a graphene contacted by two soft magnetic electrodes. The
further experiments have found that a inserted tunnel film at the
graphene/electrode interface favors MR which can reach up to
$\sim$12\%.$^{11}$ Additionally, it is expected$^{12}$ that the spin
dependence of the electron energy can be also produced by a FM
insulator substrate which acts as an effective magnetic field
parallel to the graphene layer.$^{13,14}$. Theoretically, using the
tight-binding model, Brey and Fertig$^{15}$ found that the MR is
rather small due to the weak dependence of the graphene conductivity
on the electronic parameters of FM leads. Using a continuous model
by NGF, Ding $et$ $al$.$^{16}$ have demonstrated that the TMR
exhibits a cusp around zero bias in the absence of external magnetic
field for a similar device. Recently, special attentions have been
paid to zigzag-edge graphene nanoribbons (ZGNRs) spin-valve devices
since ZGNRs exhibit fascinating phenomenon due to the size
confinement and the edge state.$^{17-19}$ For example, some
theoretical studies have predicated that ZGNR-based spin-valve
devices have a very large TMR.$^{20-25}$

On the other hand, the leads (electrodes) for spin-valve devices in
the previous works have been supposed to be two-dimensional (of
honeycomb- or square-lattice) type, but more realistically they
should be modeled as three-dimensional semi-infinite slabs. Further,
the investigation on the TMR for GNR-magnetic junctions is still
lacking so far, especially, the influence of an exchange field on
the spin-dependent transport. Another important phenomenon in
spintronics is the reverse effect to TMR, i.e., spin-polarized
electrons passing from the left FM layer into the right layer, where
the magnetization deviates the left by an angle, may exert a torque
to the right FM lead. This effect is the so-called$^{26,27}$ STT
which has been extensively studied both theoretically$^{28-32}$ and
experimentally$^{33,34}$ for conventional magnetic junction systems.
A demonstration of the spin-transfer phenomenon is the
current-induced magnetic switching, which has been confirmed
experimentally in spin-valves$^{35,36}$ and magnetic tunnel
junctions (MTJs).$^{37}$ Thus, current-induced magnetic switching
provides a powerful new tool for the study of spin transport in
magnetic nanostructures. In addition, it offers the intriguing
possibility of manipulating high-density nonvolatile magnetic-device
elements, such as magnetoresistive random access memory (MRAM),
without applying cumbersome magnetic fields.$^{38}$ However, the
previous works on spin-dependent transport through a graphene-based
system are mainly focused on the tunnel current and the TMR
effect.$^{10-16,20-25}$ The investigation on the current-induced STT
in such system is sparsely reported. Although we have addressed this
issue for a FM/graphene/FM device,$^{39}$ in this paper we extend it
to a FM/AGNR/FM system.

\begin{figure}
\center
\includegraphics[bb=170 383 455 650, width=2.5in]{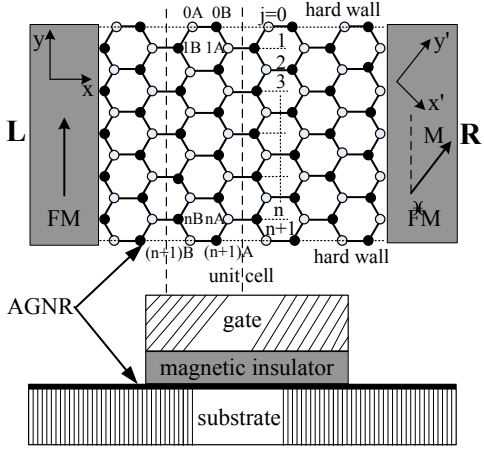}
\renewcommand{\figurename}{FIG.}
\caption {Schematic illustration of the system consisting of an AGNR
between two FM leads, where a FM insulator layer with a control gate
is deposited on AGNR. A unit cell of AGNR represented by two dashed
lines contains $n$ numbers of $A$ and $B$ sublattice sites labeled
as 1$A$, 1$B$, $\cdot\cdot\cdot$, $nA$, $nB$. Two additional hard
walls are imposed on both edges at chains $j$=0 and $n$+1. The
magnetic moment of the two leads is aligned at a relative angle
$\theta$.}
\end{figure}

In this paper, we present a theoretical investigation on the fully
spin-dependent transport through an AGNR between two FM leads with
arbitrary polarization direction at low temperatures, where a
magnetic insulator is deposited on AGNRs to induce an exchange
splitting. By using the Keldysh NGF method,$^{40,41}$ the density of
state (DOS), the linear conductance, the differential conductance,
and the current for the system have been calculated separately for
spin-up and -down channels, and consequently the TMR dependence on
applied bias has been calculated. Further, the dependence of STT on
the bias and the polarization angle of the two leads have also been
examined. It is demonstrated that, the spin-resolved transport
property of the system depends sensitively on both the widths of
AGNRs and the polarization strength of FM leads. Around zero bias
the TMR versus bias voltage possesses a pronounced plateau structure
for both semiconducting 7-AGNR and metallic 8-AGNR systems without
exchange splitting. The plateau structure for 8-AGNR system is
remarkably broader than that for 7-AGNR one. Interestingly, the
increase of the exchange splitting strength suppresses the amplitude
of this structure for 7-AGNR system. However, the TMR is enhanced at
bias that ranged from $-\Delta$ to $\Delta$ for 8-AGNR system.
Further, the current-induced STT for 7-AGNR system is systematically
larger than that for 8-AGNR system.

The rest of the paper is organized as follows. In Sec. II, we derive
the analytical expressions for the spin-dependent DOS, the
conductance, the current, and the STT, starting from the system
Hamiltonian by NGF approach. Some numerical examples and the
discussions for the results are demonstrated in Sec. III. Finally,
Sec. IV concludes the paper.

\section{Model and Method}
The geometry of the system considered in this paper is shown in Fig.
1, where an AGNR contains two sublattices denoted by $A$ and $B$. We
use $n$, the number of $A(B)$-site atoms in a unit cell, to denote
AGNR with different width. Ideal (perfect) AGNRs with $n$=3$j$-1
with positive integer $j$ are metallic, otherwise are
semiconducting.$^{18,19}$ From the top down, atoms in a unit cell
are labeled as 1$A$, 1$B$,$\cdots$, $n$$A$, $n$$B$. As shown in Fig.
1, the hard-wall condition is imposed on both edges at chains $j$=0
and $n+1$. The definition of GNRs in this work is in accord with the
previous convention,$^{19}$ i.e., an AGNR is identified by the
number of carbon zigzag chains forming the width of ribbon, and an
AGNR with $n$ carbon chains is named as $n$-AGNR, therefore its
width is $W$=$(n-1)\sqrt{3}a/2$, where $a$=1.42 ${\AA}$ is the C$-$C
bond length.

The system under consideration here is composed of a semiconducting
7-AGNR or a metallic 8-AGNR connected to two FM leads, which can be
considered as a MTJ (or a spin-valve) device. A layer of magnetic
insulator is deposited on the top of AGNR sample to induce an
exchange splitting between spin-up and -down carriers.$^{12}$ The
gate allows us to control the Fermi level locally, i.e., to create a
tunable barrier in AGNR. In this way, controlling on both charge and
spin carrier concentrations can be achieved. The magnetic moment
$M_L$ of the left FM lead is assumed to be parallel to the $y$-axis,
while the moment $M_R$ of the right FM deviates from the
$y$-direction by a relative angle $\theta$. The tunneling current
flows along the $x$-axis [see Fig. 1]. We assume that the magnitude
of exchange splitting can be modulated by the magnetic insulator. It
seems a reasonable assumption that the magnetic insulator is made of
different materials from FM leads.

The total Hamiltonian for the system considered reads
\begin{eqnarray}
H=H_{G}+H_{L}+H_{R}+H_{T},
\end{eqnarray}
where $H_G$ describes the central AGNR region, $H_{L(R)}$ is the
Hamiltonian for the left (right) FM lead, and $H_T$ for the coupling
between AGNR and leads. Here, spin-orbit interaction in graphene or
GNRs is neglected because it is too weak and is of the order of
3$-$4 meV,$^{42,43}$ it opens up a gap of the order of 10$^{-3}$ meV
at the Dirac point. In the tight-binding approximation, these
partial Hamiltonians can be respectively written as following:
\begin{eqnarray}
H_{G}=\sum_{\beta i ,\sigma}(\epsilon_0+ s\Delta)a^{\dag}_{\beta i
,\sigma}a_{\beta i,\sigma}+
t\sum_{<ij>,\sigma}(a^{\dag}_{Ai,\sigma}a_{Bj,\sigma}+\text{H.c}),
\end{eqnarray}
where $a^{\dag}_{\beta i,\sigma}$ ($a_{\beta i,\sigma}$) creates
(annihilates) an electron on site $i$ with spin $\sigma$ and
sublattice index $\beta$=$A(B)$, $\sigma$=$\uparrow$ ($\downarrow$)
represents the spin-up (-down) state of electrons, $\epsilon_0$ is
the on-site energy, $t$ ($\approx$2.75 eV) the nearest-neighbor
hopping energy, and $\Delta$ the exchange splitting energy induced
by the FM insulator on the top of AGNR, s=$\pm1$ stands for the
electron spin parallel ($+$) or antiparallel ($-$) to the exchange
field. The sum over $\langle i,j\rangle$ is restricted to the
nearest-neighbor atoms. Hamiltonian (2) can also be rewritten in
momentum space
\begin{eqnarray}
H_{G}=\sum_{\beta,\textbf{q}\sigma}(\epsilon_0+ s\Delta)
a^{\dagger}_{\beta,\textbf{q}\sigma}a_{\beta,\textbf{q}\sigma}+
\sum_{\textbf{q}\sigma}[\phi(\textbf{q})a^{\dagger}_{A,\textbf{q}\sigma}a_{B,\textbf{q}\sigma}+\text{H.c.}],
\end{eqnarray}
where
$\phi(\textbf{q})=-t[2e^{iq_xa/2}\text{cos}(\frac{\sqrt{3}a}{2}q_y)+e^{-iq_xa}]$
is the structural factor with wave-vector in the $x$-direction
within the first Brillouin zone (0$\leq$$|q_x|$$\leq$$\pi/(3a)$) and
the discretized wave-vector in the $y$-direction
$q_y$=$\frac{2}{\sqrt{3}a}\frac{m\pi}{n+1}$ ($m$=1,2, $\cdots$,
$n$).

The two Hamiltonians
\begin{eqnarray}
H_{L}=\sum_{\textbf{k},\sigma}\epsilon_{\textbf{k}L\sigma}c_{\textbf{k}L\sigma}^{\dag}c_{\textbf{k}L\sigma}
\end{eqnarray}
and
\begin{eqnarray}
H_{R}=\sum_{\textbf{k},\sigma}[\epsilon _R(\textbf{k})-\sigma
M_R\cos\theta]c_{\textbf{k}R\sigma }^{\dag}c_{\textbf{k}R\sigma
}-M_R\sin\theta c_{\textbf{k}R\sigma
}^{\dag}c_{\textbf{k}R\bar{\sigma}}
\end{eqnarray}
are respectively for the left and right lead, where
$\epsilon_{\textbf{k}L\sigma}$=$\epsilon_L(\textbf{k})-eV-\sigma
M_L$ with applied bias voltage $V$ and magnetic momentum $M_L$,
$\epsilon_{L,(R)}(\textbf{k})$ is the single-particle dispersion for
the left (right) lead,
$c^{\dag}_{\textbf{k}L(R)\sigma}(c_{\textbf{k}L(R)\sigma})$ is the
creation (annihilation) operator of an electron with wavevector
$\textbf{k}$ in the left (right) lead, $\bar{\sigma}$ denotes the
opposite spin-polarization with respect to $\sigma$. Note that here
the coupling between the left lead and AGNR involves the $B$
sublattice, while it between the right lead and AGNR involves the
$A$ sublattice (see Fig. 1). However, Hamiltonian (5) will be
slightly different if the coupling at two interfaces involves the
same sublattice, but the result is unchanged.

The coupling Hamiltonian
\begin{eqnarray}
H_{T}&=&\frac{1}{\sqrt{N}}\sum_{\textbf{k}L\textbf{q}\sigma}
(T_{\textbf{k}L\textbf{q}}c_{\textbf{k}L\sigma}^{\dag}a_{A,\textbf{q}\sigma
}+\text{H.c.})\nonumber\\
&&+\frac{1}{\sqrt{N}}\sum_{\textbf{k}R\textbf{q}\sigma}
(T_{\textbf{k}R\textbf{q}}c_{\textbf{k}R\sigma}^{\dag}a_{B,\textbf{q}\sigma
}+\text{H.c.}),
\end{eqnarray}
where $T_{\textbf{k}L\textbf{q}}$ ($T_{\textbf{k}R\textbf{q}}$) is
the coupling matrix between left (right) lead and AGNR, $N$ is the
number of sites in sublattice $A$ or $B$. It should be pointed out
that the annihilation operator $a_{B,\textbf{q}\sigma}$ in Eq. (6)
should be replaced by $a_{A,\textbf{q}\sigma}$ if the coupling
between right lead and AGNR involves the $A$ sublattice.

Now we employ the usually defined retarded, advanced and correlated
Green's function (GF)$^{40,41}$
$G_{\textbf{q}a,\textbf{q}'a}^{\sigma\sigma', r}(t_2,t_1)=
-i\theta(t_2-t_1)\langle{\{a_{\textbf{q}\sigma}(t_2),a_{\textbf{q}'\sigma'}^\dag{(t_1)}}\}\rangle,$
$G_{\textbf{q}a,\textbf{q}'a}^{\sigma\sigma',a}(t_2,t_1)=
i\theta(-t_2$+$ t_1)
\langle{\{a_{\textbf{q}\sigma}(t_2),a_{\textbf{q}'\sigma'}^\dag{(t_1)}}\}\rangle$,
and $G_{\textbf{q}a,\textbf{q}'a}^{\sigma\sigma',<}(t_2,t_1)=
i\langle{a_{\textbf{q}'\sigma'}^\dag(t_1),a_{\textbf{q}\sigma}{(t_2)}}\rangle$
to the total Hamiltonian in spin space. When the unperturbed GF
$g^{r,a}_{\textbf{q}a,\textbf{q}a}(\omega)$ of the AGNR is obtained,
one can obtain $G^{r,a}_{\textbf{q}a,\textbf{q}'a}(\omega)$ and
$G^<_{\textbf{q}a,\textbf{q}'a}(\omega)$ from the standard GF
technique and the Keldysh equation$^{40}$
\begin{eqnarray}
G^{r,a}_{\textbf{q}a,\textbf{q}'a}(\omega)=\delta_{\textbf{q}\textbf{q}'}g^{r,a}_{\textbf{q}a,
\textbf{q}a}(\omega)+g^{r,a}_{\textbf{q}a,\textbf{q}a}(\omega)T^{r,a}(\omega)g^{r,a}_{\textbf{q}'a,\textbf{q}'a}(\omega),
\end{eqnarray}
\begin{eqnarray}
G^<_{\textbf{q}a,\textbf{q}'a}(\omega)=G^r_{\textbf{q}a,\textbf{q}'a}(\omega)\Sigma^<(\omega)
G^a_{\textbf{q}a,\textbf{q}'a}(\omega),
\end{eqnarray}
where
$T^{r,a}(\omega)=\Sigma^{r,a}(\omega)/[1$$-$$g^{r,a}_{aa}(\omega)\Sigma^{r,a}(\omega)]$
with
$g^{r,a}_{aa}(\omega)=1/N{\sum_\textbf{q}}g^{r,a}_{\textbf{q}a,\textbf{q}a}(\omega)$
and $\Sigma^{r,a}(\omega)=\mp
i/2[\bm{\Gamma}_{L\textbf{q}\textbf{q}'}(\omega)+R\bm{\Gamma}_{R\textbf{q}\textbf{q}'}R^\dag]$
is the retarded/advanced self-energy, and
$\Sigma^<(\omega)=i[f_L(\omega)\bm{\Gamma}_{L\textbf{q}\textbf{q}'}(\omega)
+f_R(\omega)R\bm{\Gamma}_{R\textbf{q}\textbf{q}'}R^\dag]$ is the
correlated self-energy. In the above equations, the matrix
\begin{eqnarray*}
R=\left(
           \begin{array}{cc}
             \text{cos}(\theta/2) & -\text{sin}(\theta/2) \\
             \text{sin}(\theta/2) & \text{cos}(\theta/2) \\
           \end{array}
         \right),
\end{eqnarray*}
$f_{L(R)}(\omega)$=$1/[{e^{(\omega-\mu_{L(R)})/k_BT}+1}]$ is the
Fermi-Dirac distribution function of the left (right) FM lead, the
linewidth function
\begin{eqnarray*}
\bm{\Gamma}_{\alpha \textbf{q}\textbf{q}'}(\omega)=\left(
           \begin{array}{cc}
             \bm{\Gamma}_{\alpha \textbf{q}\textbf{q}'\uparrow}(\omega) & 0 \\
             0 & \bm{\Gamma}_{\alpha \textbf{q}\textbf{q}'\downarrow}(\omega) \\
           \end{array}
         \right)
\end{eqnarray*}
with $\bm{\Gamma}^{\sigma}_{\alpha
\textbf{q}\textbf{q}'}(\omega)$=$2\pi\sum_{\textbf{k}}T^*_{\textbf{k}\alpha
\textbf{q}}T_{\textbf{k}\alpha
\textbf{q}'}\delta(\omega-\varepsilon_{\textbf{k}\alpha\sigma})$
describes the influence of the leads. Here, we consider electrons
near the Fermi level $E_F$ which contribute predominantly to the
tunneling. In this case one may assume that the coupling matrix is
independent of energy and set $\bm{\Gamma}^{\sigma}_{\alpha
\textbf{q}\textbf{q}'}$=$\bm{\Gamma}^{\sigma}_{\alpha}$.

Furthermore, the isolated retarded GF for an AGNR is defined as
\begin{eqnarray}
g^{r}_{aa}(\omega)=\sum_{q_x,q_y,\pm}\frac{|\psi\rangle_{\pm\pm}\langle\psi|}{[\omega-(\epsilon_{0}+s\Delta)]-E_{\pm}+i\eta^+}
\end{eqnarray}
with a positive infinitesimal $\eta^+$, the dispersion relation
$E_{\pm}$=$\pm|\phi(\textbf{q})|$ and the wave function
$|\psi\rangle_{\pm}$=$\frac{\sqrt{2}}{2}(|\psi\rangle_A\pm\sqrt{\frac{\phi^*(\textbf{q})}{\phi(\textbf{q})}}|\psi\rangle_B)$,
where the plus (minus) sign applies to the upper (lower) $\pi^*$
($\pi$) band,
$|\psi\rangle_A$=$\sqrt{\frac{2}{N_x(n+1)}}\sum^n_{m'=1}\sum_{x_{A_{m'}}}e^{iq_xx_{A_{m'}}}\text{sin}(\frac{\sqrt{3}q_ya}{2}m')|A_{m'}\rangle$
and
$|\psi\rangle_B$=$\sqrt{\frac{2}{N_x(n+1)}}\sum^n_{m'=1}\sum_{x_{B_{m'}}}e^{iq_xx_{B_{m'}}}\text{sin}(\frac{\sqrt{3}q_ya}{2}m')|B_{m'}\rangle$
with the number of unit cells $N_x$ along the $x$-direction.$^{19}$
A direct calculation yields expressions for the GF of an AGNR
\begin{widetext}
\begin{eqnarray}
g^{r}_{aa}(\omega)&=&\frac{12a}{N_x(n+1)}\frac{\omega-(\epsilon_{0}+s\Delta)+i\eta^+}{2\pi}\int^{\frac{\pi}{3a}}_{\frac{-\pi}{3a}}dq_x
\sum^n_{m,m'=1}\frac{\text{sin}^2(\frac{m\pi}{n+1})m'}{[\omega-(\epsilon_{0}+s\Delta)+i\eta^+]^2-t^2[1+4\text{cos}^2\frac{m\pi}{n+1}+
4\text{cos}\frac{m\pi}{n+1}\text{cos}(3q_x/2)]}.
\end{eqnarray}
\end{widetext}

The exchange term splits the system into two separate spin
subsystems. Therefore, the DOS, the linear conductance, and the
current for spin-up and -down channels can be respectively obtained
as
\begin{eqnarray}
\rho_{\uparrow(\downarrow)}(\omega)=-\frac{1}{\pi}\text{Im}
G_{a\uparrow\uparrow(\downarrow\downarrow)}^r(\omega ),
\end{eqnarray}
\begin{eqnarray}
G_{\uparrow(\downarrow )}=\frac{e^{2}}{h} X_{\uparrow \uparrow
(\downarrow \downarrow )}(\omega)|_{\omega=E_F},
\end{eqnarray}
\begin{eqnarray}
I_{\uparrow (\downarrow )}(V)=\frac{e}{\hbar}\int_{-eV/2}^{eV/2}
\frac{d\omega }{2\pi}(f_{R}-f_{L})X_{\uparrow \uparrow (\downarrow
\downarrow)}(\omega),
\end{eqnarray}
where $V$ is the applied bias voltage, and the matrix elements in
Eqs. (11)-(13) are defined as
\begin{eqnarray}
G_{a}^{r}(\omega)=\sum_{\textbf{q}\textbf{q}^{\prime}}
G_{\textbf{q}a,\textbf{q}^{\prime }a}^{r(a)}(\omega )=\left(
           \begin{array}{cc}
             G_{a\uparrow\uparrow}^{r}(\omega ) & G_{a\uparrow\downarrow}^{r}(\omega ) \\
             G_{a\downarrow\uparrow}^{r}(\omega ) & G_{a\downarrow\downarrow}^{r}(\omega ) \\
           \end{array}
         \right)
\end{eqnarray}
and
\begin{eqnarray}
X(\omega)=[G_a^{r}(\omega)(R\bm{\Gamma}
_{R}R^{\dag})]G_a^a(\omega)\bm{\Gamma}_{L}=\left(
           \begin{array}{cc}
             X_{\uparrow\uparrow}(\omega) & X_{\uparrow\downarrow}(\omega) \\
             X_{\uparrow\downarrow}(\omega) & X_{\downarrow\downarrow}(\omega) \\
           \end{array}
         \right).
\end{eqnarray}

On the other hand, the STT is the time evolution rate of the total
spin of left or right FM lead.$^{26,27}$  Here, we ignored the
interlayer exchange coupling, so the out-of-plane torque is
zero.$^{44}$ By means of the NGF method, the current-induced
in-plane STT along $x'$-direction in the $(x',y')$ coordinate frame
[see Fig. 1] exerting on the right FM lead can be obtained$^{32}$
\begin{eqnarray}
\tau_R^{x^{\prime}}&=&\frac{1}{4\pi }\int d\omega (f_{R}-f_{L})\text{Tr}[%
\text{G}_{a}^{r}(\omega )\Gamma_{L}\text{G}_{a}^{a}(\omega )R\Gamma
_{R}R^{\dag}\nonumber\\&&\cdot(-\cos \theta \sigma _x+\sin \theta
\sigma _z)],
\end{eqnarray}
where $\sigma_x$ and $\sigma_z$ are the Pauli matrices.

\section{Results and Discussions}
In what follows, we present some numerical examples of
$\rho_{\uparrow (\downarrow )}$, $G_{\uparrow (\downarrow )}$,
$G_{d,\uparrow (\downarrow )}$, TMR, and $\tau_R^{x^{\prime}}$ for
the system with 7- or 8-AGNR at low temperatures according to Eqs.
(11)-(16). In the calculation, the two FM leads are assumed to be
made of the same materials, i.e., $p_L$=$p_R$=$p$, where
$p_{L(R)}$=$(\Gamma_{L(R)\uparrow}-\Gamma_{L(R)\downarrow})/
(\Gamma_{L(R)\uparrow}+\Gamma_{L(R)\downarrow})$ is the polarization
strength of the left (right) FM. Under the wide bandwidth
approximation,
$\Gamma_{L\uparrow(\downarrow)}$=$\Gamma_{R\uparrow(\downarrow)}$=$\Gamma_0(1\pm
p)$. The TMR is conventionally defined as
$\text{TMR}$=$(I_P-I_{AP})/I_P$, where $I_P$ and $I_{AP}$ are the
total current for the parallel and antiparallel configuration,
respectively. The coupling $\Gamma_0$  between the AGNR and leads
without internal magnetization is taken as the energy unit. The
chemical potentials for the two FM leads are set as
$\mu_{L,R}$=$E_F\pm$0.5 eV with the Fermi energy $E_{F}$=0. The
on-site energy $\epsilon_0$=0, and $\Gamma_0$=$t$ because that they
are should be in the same order.$^{45}$

\subsection{Spin-dependent transport}
\begin{figure}
\center
\includegraphics[bb=24 54 584 504,width=8.5cm]{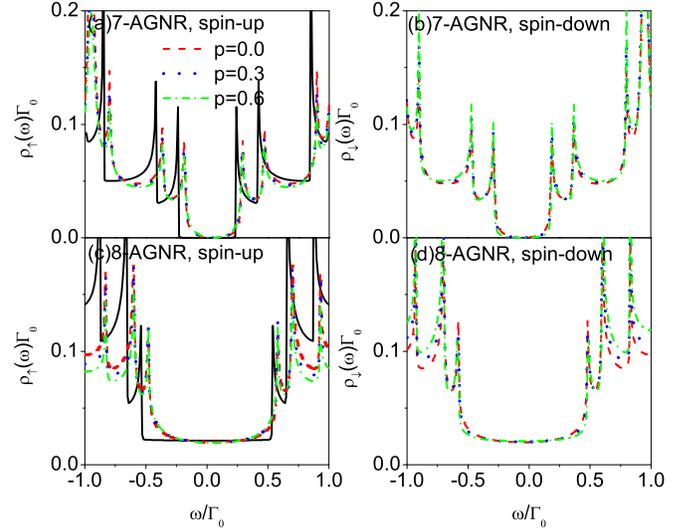}
\renewcommand{\figurename}{FIG.}
\caption{(Color online) DOS for spin-up [(a) and (c)], and -down
[(b) and (d)] channels as a function of energy $\omega$ (in units of
$\Gamma_0$) with different polarization $p$, where the parameters
are taken as $\epsilon_0$=0, $\theta$=0 and $\Delta$=$0.05$. For
comparison, the DOS for bare ($\Delta$=$\Gamma_0$=$0$) 7-AGNR (a)
and 8-AGNR (c) is shown in (dark) solid lines.}
\end{figure}

In Fig. 2, we present the spin-dependent DOS versus energy $\omega$
(in units of $\Gamma_0$) in parallel configuration ($\theta$=0) with
exchange splitting $\Delta$=0.05 and different polarization strength
$p$ for the system with 7- and 8-AGNR, respectively. In the case of
bare AGNR ($\Delta$=$\Gamma$=$0$), a zero value plateau in DOS for
7-AGNR appears symmetrically with respect to the Fermi level [see
(dark) solid line in Fig. 2(a)] due to the presence of the energy
gap for semiconducting 7-AGNR, while for 8-AGNR a nonzero
symmetrical plateau with respect to the Fermi level appears [see
(dark) solid line in Fig. 2(c)] because of the gapless energy band
in metallic 8-AGNR. However, when a magnetic deposition is applied,
the positions of plateaus for both 7-AGNR and 8-AGNR systems shift
with the exchange field strength $\Delta$: the spin-up (-down) DOS
deviates the origin point by the magnitude of $\Delta$ towards the
positive (negative) direction of the energy. The reason for this
phenomenon is that the exchange interaction in the magnetic
deposition behaves as an effective in-plane magnetic field, and acts
on the electrons in AGNRs, thus leading to the shift of the energy
level related to the electron spin. For larger $\omega$,
$\rho_{\sigma}$ shows an oscillation behavior with sharp peaks,
which indicates the buildup of Van Hove singularities (VHSs) at
subband edge. Additionally, we notice that as $p$ increases from
zero [(red) dashed lines for normal leads] to 0.6 [(green)
dash-dotted lines] in Fig. 2, both 7- and 8-AGNR systems are almost
immune from $p$, which is different from the bulk graphene tunneling
junction.$^{39}$ This stems from the fact the VHSs induced by the
edge effect of AGNRs are independent on $p$.

\begin{figure}
\center
\includegraphics[bb=24 54 584 504,width=8.5cm]{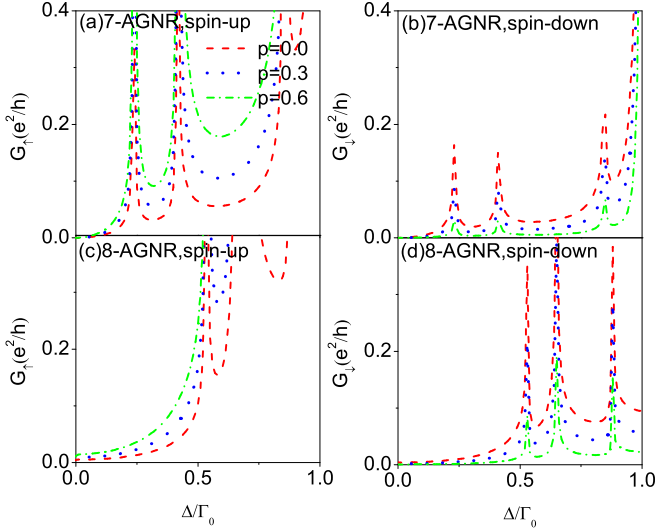}
\renewcommand{\figurename}{FIG.}
\caption{(Color online) The linear conductance as a function of
exchange splitting $\Delta$ with different polarization $p$ for
spin-up [(a) and (c)], and spin-down [(b) and (d)] channels. The
other parameters are the same as those of Fig. 2.}
\end{figure}

Figure 3 shows the spin-dependent linear conductance (in units of
$e^2/h$) versus exchange splitting $\Delta$ (in units of $\Gamma_0$)
with different polarization $p$ in parallel alignment at the Fermi
energy for the system with 7- and 8-AGNR, respectively. It is seen
that for both 7-AGNR and 8-AGNR systems, $G_{\uparrow}$ increases
[see Figs. 3(a) and 3(c)] while $G_{\downarrow}$ decreases [see
Figs. 3(b) and 3(d)] with the increase of polarization $p$, which is
due to the conventional spin-valve effect in the magnetic tunneling
junction. Namely, with increasing $p$, the proportion of spin-up
electrons increases and that of spin-down electrons decreases in
AGNRs. Consequently, it becomes easier for spin-up electrons to
tunnel through the barrier, but harder for spin-down ones due to the
presence of the inverse spin-direction for spin-down electrons in
the tunneling process, thus we conclude that $G_{\uparrow}$
increases while $G_{\downarrow}$ decreases with the increase of $p$.
In addition, it is found that for the small $\Delta$ (below the full
spin polarization$^{46}$), the conductance value is zero for 7-AGNR
system, while nonzero constant for 8-AGNR system. This is related to
the presence (absence) of energy gap in semiconducting 7-AGNR
(metallic 8-AGNR). For a larger $\Delta$ (above the full spin
polarization), the conductance for either 7- or 8-AGNR system
displays an oscillation enhancement behavior with shape peaks, which
stems from the resonant tunneling through different subbands due to
the edge effect.$^{47,48}$

\begin{figure}
\center
\includegraphics[bb=24 54 584 504,width=8.5cm]{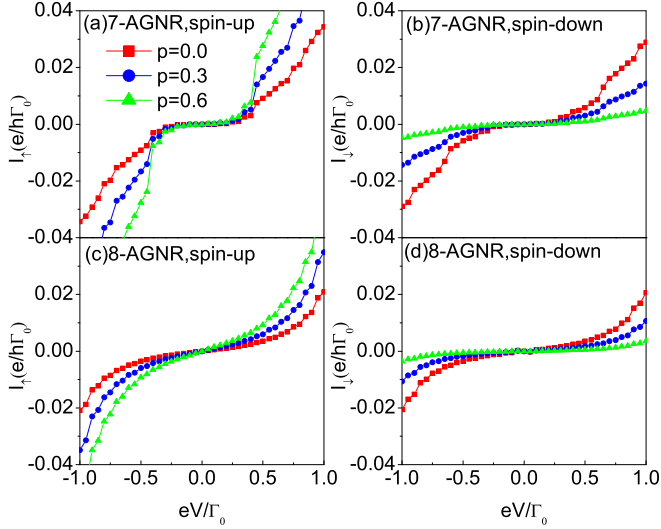}
\renewcommand{\figurename}{FIG.}
\caption{(Color online) The I-V curve with different polarization
$p$ for spin-up [(a) and (c)], and spin-down [(b) and (d)] channels.
The parameters are the same as those of Fig. 2.}
\end{figure}

The bias voltage dependence of current (in units of $e/h\Gamma_0$)
for the 7-AGNR and 8-AGNR systems with different polarization $p$
under the parallel configuration is demonstrated in Fig. 4. It is
clearly seen that the current has a step-like structure as a
function of bias voltage, which results from the constant DOS at the
Fermi energy and resonance due to VHSs. In particular, the
dependence of the current on the polarization $p$ exhibits different
behaviors for the 7-AGNR and 8-AGNR systems. With increasing $p$,
the current increases in the entire bias voltage for the 8-AGNR
system, while for the 7-AGNR system, the current rises only in the
large bias, and almost remains zero at small bias voltage. The
explanation for this phenomenon is as follows: the DOS for the
7-AGNR system vanishes near the Fermi level, thus the 7-AGNR system
resembles a insulator-like barrier at the low energy. In this case,
the electrons will difficultly tunnel through the 7-AGNR system at
low bias.

\begin{figure}
\center
\includegraphics[bb=24 54 584 504,width=8.5cm]{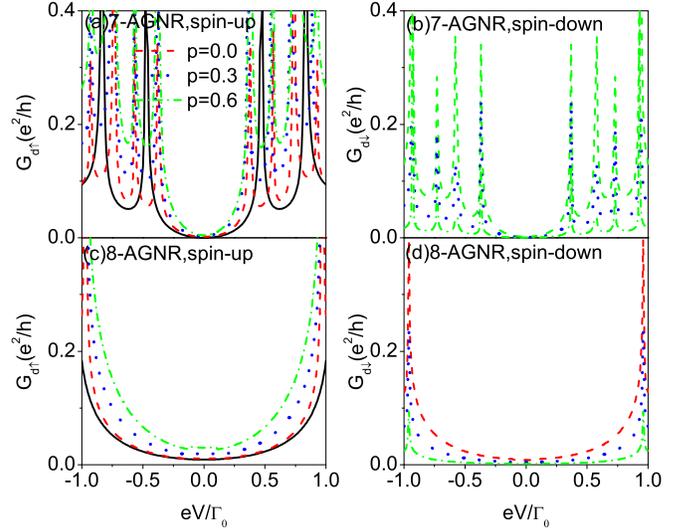}
\renewcommand{\figurename}{FIG.}
\caption{(Color online) The differential conductance as a function
of bias $V$ with different polarization $p$ for spin-up [(a) and
(c)], and spin-down [(b) and (d)] channels. The other parameters are
the same as those of Fig. 2. For comparison, the differential
conductance for 7- and 8-AGNR with $\Delta$=$p$=0 is also shown in
(dark) solid lines.}
\end{figure}

In Fig. 5, we demonstrate the dependence of differential conductance
$G_d$=$dI/dV$ (in units of $e^2/h$) on the bias voltage for the
system with different polarization $p$ under the parallel
configuration. For comparison, we plot the differential conductance
of the 7- and 8-AGNR systems for $\Delta$=$p$=$0$ in (dark) solid
lines. The differential conductance exhibits the successive
oscillation peaks corresponding to the resonant tunneling through
the edge-induced subbands. In addition, we find that there exists a
respective plateau structure at small bias for the 7- and 8-AGNR
systems. In the range of the plateau, the conductance approaches to
zero for the 7-AGNR system, while a finite constant for 8-AGNR
system. An interesting characteristic is that the plateau structure
for the 8-AGNR system is broader than that for the 7-AGNR system
since 8-AGNR has a larger interval between the Fermi energy and the
lowest subband [see Fig. 2]. When $\Delta$ is applied, the peak
splits into two peaks located at the two sides of the original peak
for both 7-AGNR and 8-AGNR systems, and the position of the peaks
depends on the magnitude of the exchange splitting. This behavior is
due to the exchange splitting of the edge-induced subbands. From
Fig. 5, $G_{d,\uparrow}$ increases while $G_{d,\downarrow}$
decreases for both 7- and 8-AGNR systems with increasing $p$, which
is a typical spin-valve effect.

\begin{figure}
\center
\includegraphics[bb=55 50 393 512,width=6cm]{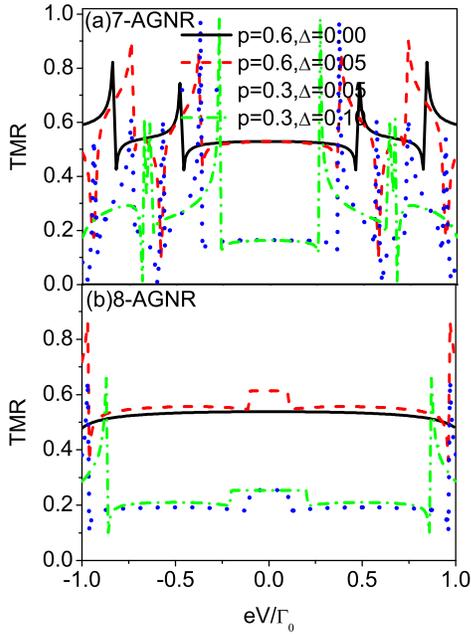}
\renewcommand{\figurename}{FIG.}
\caption{(Color online) The dependence of TMR on bias $V$ for (a)
7-AGNR, (b) 8-AGNR system with different polarization $p$ and
splitting $\Delta$.}
\end{figure}

Figure 6 shows the TMR versus the bias voltage $V$ for 7- and 8-AGNR
systems with different polarization $p$ and exchange splitting
$\Delta$. Firstly, it is noted that the TMR for the 7- and 8-AGNR
systems has a high value relative to the conventional magnetic
junction; a pronounced plateau for 7- and 8-AGNR systems appears at
lower bias due to the constant DOS near the Fermi energy [see Fig.
2]. These results are quite consistent with the previous
studies.$^{39,44}$ For high bias, the TMR exhibits the successive
oscillation peaks resided near the edge-induced subbands. The
amplitude of these peaks can even approach to 1 indicating an
perfect spin-valve effect. A high TMR for the system with an AGNR
has also been obtained in Refs. [49,50]. It should be pointed out
that the result here is different from that of Ref. [23] where the
TMR for the AGNRs systems is very small in the vicinity of the Fermi
energy. The plateau structure for 8-AGNR system is systematically
broader than that corresponding to 7-AGNR system, which may be very
useful to overcome a well-known shortcoming of MRAM, which is the
large decrease in TMR with applied bias voltage. Interestingly, the
increase of $\Delta$ suppresses the amplitude of this structure for
7-AGNR system. However, the TMR is enhanced within bias range from
-$\Delta$ to $\Delta$ for 8-AGNR system. This is caused by the
exchange splitting influence on the DOS as mentioned in Fig. 2. The
results are quantitatively different from our previous work for
FM/graphene/FM system due to the finite size for AGNRs.$^{16,39}$
When the polarization changes to $p$=0.3, as shown by the (blue)
dotted lines and (green) dash-dotted lines in Fig. 6, the TMR for
both 7-AGNR and 8-AGNR systems becomes obviously smaller because the
leads become less spin-polarized with decreasing $p$.$^{51}$ This
result is different from Ref. [15], where the TMR is nearly
independent of the electronic details of the leads. We speculate
that the structure in the Ref. [15] is very different from the
structure considered here. Our results suggest that it is more
favorable to fabricate a graphene-based spin-valve device by using a
metallic AGNR.

\subsection{Spin transfer torque}
\begin{figure}
\center
\includegraphics[bb=28 27 327 540,width=7cm]{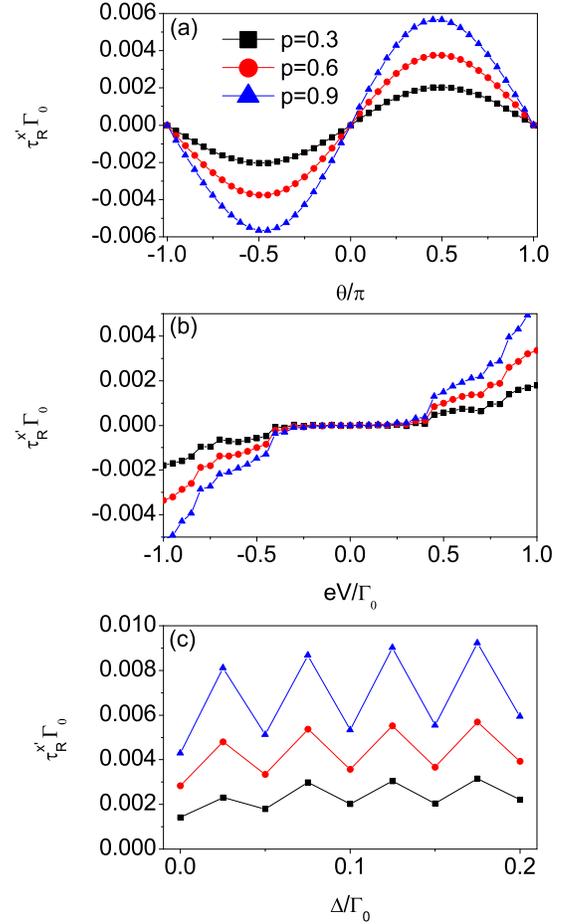}
\renewcommand{\figurename}{FIG.}
\caption{(Color online) The dependence of current-induced STT for
7-AGNR on (a) relative orientation angle $\theta$ at the bias
voltage $V$=$0.5\Gamma_0/e$, (b) bias voltage $V$ at
$\theta$=$\pi/3$, and (c) exchange splitting $\Delta$ at
$\theta$=$\pi/3$ and $V$=$0.5\Gamma_0/e$ with different polarization
$p$. The other parameters are the same as in Fig. 2.}
\end{figure}
\begin{figure}
\center
\includegraphics[bb=28 27 327 540,width=6.5cm]{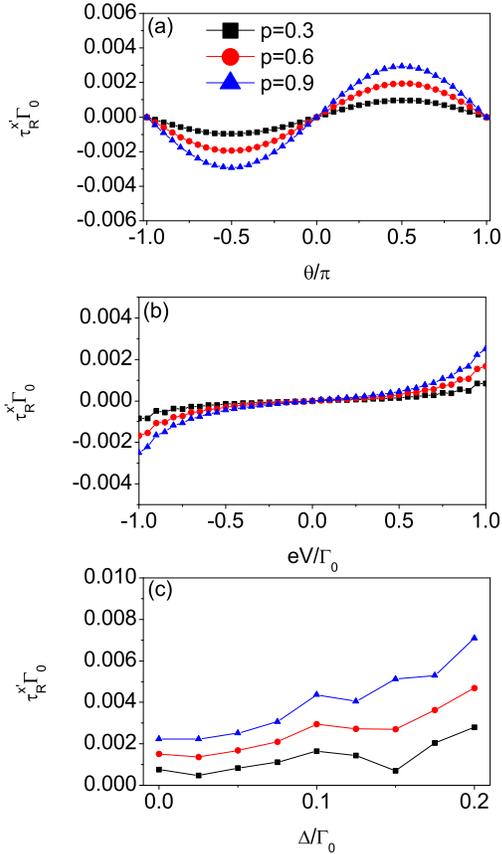}
\renewcommand{\figurename}{FIG.}
\caption{(Color online) The dependence of current-induced STT for
8-AGNR on (a) relative orientation angle $\theta$ at the bias
voltage $V$=$0.5\Gamma_0/e$, (b) bias voltage $V$ at
$\theta$=$\pi/3$, and (c) exchange splitting $\Delta$ at
$\theta$=$\pi/3$ and $V$=$0.5\Gamma_0/e$ with different polarization
$p$. The other parameters are the same as in Fig. 2.}
\end{figure}

Figures 7 and 8 show the current-induced STT ($\tau_R^{x^{\prime}}$)
as functions of angular $\theta$, bias voltage and exchange
splitting strength for 7-AGNR and 8-AGNR systems with different
polarization $p$. It is seen that $\tau_R^{x^{\prime}}$ versus the
angle $\theta$ shows a sine-like behavior for 7-AGNR and 8-AGNR
systems, which is in line with the previous findings.$^{28,39}$ A
similar result has also been obtained for a FM/NM/FM trilayer system
discussed in Ref. [52], although the transport mechanisms are
different. The present result can be easily understood because the
spin torque is proportional to $\mathbf{S_R\times(S_L\times S_R)}$,
where $\mathbf{S_L}$ and $\mathbf{S_R}$ are the spin moments of the
left and right FMs, respectively. Thus we can conclude that
$\tau_R^{x^{\prime}}$ vanishes when the relative alignment of
magnetization of the two FM leads is parallel ($\theta$=0) or
antiparallel ($\theta$=$\pi$). Furthermore, the STT for 7-AGNR
system is quantitatively larger than that for 8-AGNR system,
suggesting that a semiconducting AGNR is a better choice for a
current-controlled magnetic memory element.

In addition, it is evident that with increasing polarization $p$,
the STT is enhanced, as displayed in Fig. 7(a) and Fig. 8(a). This
is in line with the statement that $\tau_R^{x^{\prime}}$ is
proportional to the polarization strength of FM.$^{28}$ Similar
result is obtained in the conventional spin-valve device.$^{26}$
From Figs. 7(b) and 8(b), one can notice the step-like feature in
the STT as a function of bias voltage for both 7- and 8-AGNR
systems, which is reminiscent of the current versus the bias. The
reason is that the in-plane $\tau_R^{x^{\prime}}$ is proportional to
the difference of spin-polarized current
($I^s(0)-I^s(\pi))$,$^{28,31}$ where
$I^s(\theta)=I_{\uparrow}(\theta)-I_{\downarrow}(\theta)$ is the
spin-current densities for the angle $\theta$. In practice,
$I^s(\pi)$ (not shown here) for both 7- and 8-AGNR systems vanishes
identically due to the absence of minority states available for
tunneling. Thereby, the bias dependence of $\tau_R^{x^{\prime}}$
remains the similar feature to that of $I^s(0)$. This is also the
reason why the anomalous bias dependence of the STT has not been
predicted in this paper according to Ref. [31], where anomalous bias
dependence of the STT only appears in the case of $I^s(0)$$\leq
$$I^s(\pi)$. Another reason for this is that the result is obtained
here under the wide bandwidth approximation. From Figs. 7(c) and
8(c), it can be found that $\tau_R^{x^{\prime}}$ with the exchange
splitting $\Delta$ displays a oscillation behavior for the 7-AGNR
and 8-AGNR systems. This result stems from the fact that the
exchange splitting shifts the VHSs [cf. Figs. 2(a) and (c)], thus
leads to the edge-induced subband crossing of the bias windows
causing the tunneling resonance.

\section{Summary and Conclusion}
In conclusion, we have demonstrated the effects of the AGNR width,
the polarization strength of FM leads with arbitrary polarization
directions and the exchange splitting on the spin-resolved transport
properties for both semiconducting and metallic AGNRs. In contrast
to other related theoretical works, the leads in our system are not
supposed to be two-dimensional of honeycomb- or square-lattice type,
but is more realistical modeled as three-dimensional semi-infinite
FM slabs. By means of Keldysh NGF method, it is found that the
spin-resolved transport property for the system depends sensitively
on both the AGNR type and the polarization strength of FM leads.
Around zero bias the TMR versus bias voltage possesses a pronounced
plateau structure for both semiconducting 7-AGNR and metallic 8-AGNR
systems without exchange splitting. This phenomenon may be very
useful to overcome a well-known shortcoming of MRAM, which is the
large decrease in TMR with applied bias voltage. Remarkably, the
plateau structure for 8-AGNR system is much broader than that for
7-AGNR system. Interestingly, the increase of the exchange splitting
$\Delta$ suppresses the amplitude of this structure for 7-AGNR
system. However, the TMR is enhanced at bias that ranged from
-$\Delta$ to $\Delta$ for 8-AGNR system. So it may be useful in the
design of spin-valve device and graphene nanoribbons-based
spintronic devices by using a metallic AGNR. In addition, the
current-induced STT for 7-AGNR system is systematically larger that
that for 8-AGNR system, which may be more favorable to fabricate a
current-controlled magnetic memory element by using a semiconducting
AGNR. As a development of graphene fabrication technology, several
tens of nanometer wide GNRs can be obtained by patterning graphene
into a narrow ribbon, and the large exchange splitting can also be
realizable in GNRs, thus the model considered in this paper may have
potential application in spintronic devices.

\begin{acknowledgments}
This work was supported by the National Natural Science Foundation
of China (Grant Nos. 10974052 and 10574042), the Program for
Changjiang Scholars and Innovative Research Team in University
(PCSIRT, No. IRT0964), and the Scientific Research Fund of Hunan
Provincial Education Department (Grant No.09B079).
\end{acknowledgments}



\begin{references}
\bibitem{46}The full spin polarization, namely, the graphene fully ferromagnetic
depends on the magnitude of the exchange splitting, which depends on
the graphene carrier density (Ref. 13). It is estimated that the
values of the exchange splitting to make the graphene fully
ferromagnetic are in the range of 0.05$-$0.2 eV when the carrier
density is in the range of 10$^{11}$$-$10$^{12}$ (Refs. 12 and 13).
\end{references}
\end{document}